\documentclass[twoside,showpacs,superscriptaddress,twocolumn,floatfix,a4paper,aps]{revtex4}

\usepackage{color}
\usepackage{graphicx}
\usepackage[utf8x]{inputenc}
\usepackage[T1]{fontenc}
\usepackage{siunitx}
\usepackage{amstext}
\usepackage{textgreek}

\begin{document}

\title{Luminescence-induced noise in single photon sources based on BBO crystals}

\author{Radek Machulka}
\email{radek.machulka@upol.cz}
\affiliation{RCPTM, Joint Laboratory of Optics of Palacký University and Institute of Physics of Academy of Sciences of the Czech Republic, 17. listopadu 12, 771 46 Olomouc, Czech Republic}

\author{Karel Lemr}
\email{k.lemr@upol.cz}
\affiliation{Institute of Physics of Academy of Sciences of the Czech Republic, Joint Laboratory of Optics of PU and IP AS CR,
   17. listopadu 50A, 772 07 Olomouc, Czech Republic}

\author{Ondřej Haderka}
\email{ondrej.haderka@upol.cz}
\affiliation{Institute of Physics of Academy of Sciences of the Czech Republic, Joint Laboratory of Optics of PU and IP AS CR,
   17. listopadu 50A, 772 07 Olomouc, Czech Republic}

\author{Marco Lamperti}
\affiliation{Dipartimento di Scienza e Alta Tecnologia, Universit\`a degli Studi dell'Insubria and CNISM UdR Como, Via Valleggio 11, I-22100 Como, Italy}

\author{Alessia Allevi}
\affiliation{Dipartimento di Scienza e Alta Tecnologia, Universit\`a degli Studi dell'Insubria and CNISM UdR Como, Via Valleggio 11, I-22100 Como, Italy}

\author{Maria Bondani}
\affiliation{Istituto di Fotonica e Nanotecnologie, CNR, and CNISM UdR Como, Via Valleggio 11, I-22100 Como, Italy}

\date{\today}

\begin{abstract}
Single-photon sources based on the process of spontaneous parametric down-conversion play a key role in various applied disciplines of quantum optics. We characterize intrinsic luminescence of BBO crystals as a source of non-removable noise in quantum-optics experiments. By analysing its spectral and temporal properties together with its intensity, we evaluate the impact of luminescence on single-photon state preparation using spontaneous parametric down-conversion.
\end{abstract}


\maketitle

\section{Introduction}
Single-photon sources are an indispensable ingredient for many quantum-optics experiments. In quantum computing and communications, for instance, single photons are frequently used as quantum information carriers \cite{Nielsen_QCQI,Zeilinger_QIP}. Qubits can be conveniently encoded into their polarization \cite{Halenkova2012detector}, spatial modes \cite{Bartuskova2007cloning} or orbital angular momentum \cite{Nagali2010cloning}. More complex entangled multi-photon states can be synthesised from single photons by techniques of quantum-state engineering \cite{bib:fiurasek:multimode_fockstates,bib:zavatta:single_add,bib:mitchell:super,Lemr2008engineering}. Engineered states are particularly useful also in other fields of quantum optics such as quantum metrology \cite{Giovannetti2011}, quantum lithography \cite{Boto2000} or ghost imaging \cite{Pittman1995}.

Single photons can be generated in a number of physical processes including spontaneous parametric down-conversion (SPDC) \cite{Mandel1995}, quantum dot luminescence \cite{Michler2000} or semiconductor structure emission \cite{He2013semiconductor}. For its experimental feasibility SPDC is currently the most frequent and popular technique, at least in proof-of-principle experiments \cite{bib:bouwmeester:teleport,Kiesel2005cphase,OBrien2007comp,Lemr2011cphase,Micuda2012ampl,Vitelli2013fusion}. This method makes use of a non-linear optical medium to probabilistically transform a fraction of a strong pump beam into pairs of correlated output photons denoted as signal and idler. Photons generated in the process of SPDC originate randomly, but always in pairs. Consequently there are strong temporal \cite{Hong1987} and spatial \cite{Hamar2010} correlations between signal and idler photons of the same pair. Specific geometric configurations of this process yield two undistinguishable photons suitable for a large number of applications including cryptography \cite{BB84,E91,gisin2002crypto,R04} and quantum gates \cite{Lemr2011cphase,Lemr2012multi,Micuda2013gate,Bartkiewicz2013attack}. The existence of strong temporal correlation between the photons makes it possible the detection of one of them to herald the presence of the other \cite{Burnham1970source,Chen2006source,Walther2007source,Hunault2010source}. Using photon-number-resolving detection, heralding of multi-photon nonclassical states with suppressed photon-number fluctuations is also feasible \cite{perina2013herald,lamperti2014herald}. This way one can at least partially compensate for the probabilistic nature of the SPDC.

\begin{figure}
\includegraphics[width=8.5cm]{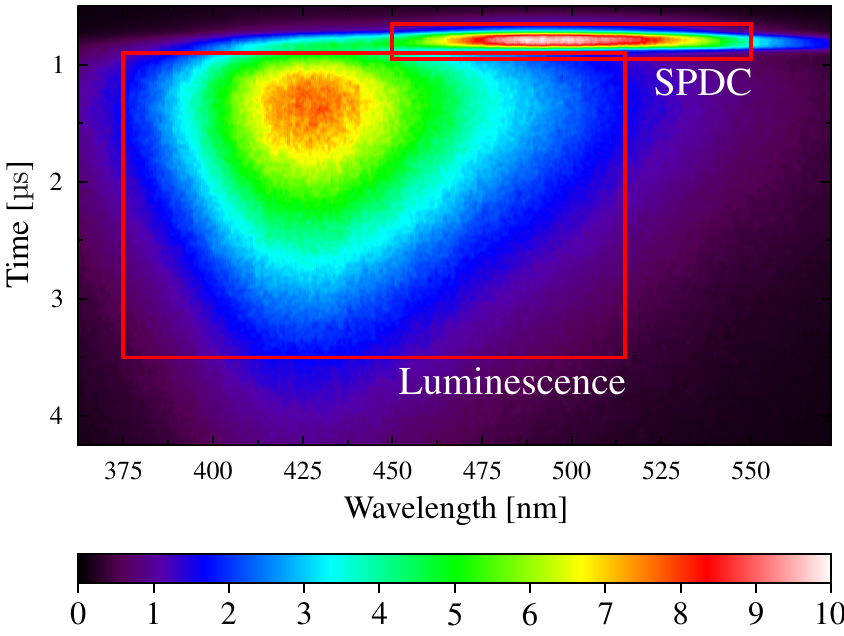}
\caption{\label{fig:streak} (color online) Typical recording of spectrally and time-resolved image obtained with a streak camera in the \si{\micro\second} time window.}
\end{figure}
Single-photon sources based on SPDC are however burdened by an inherent source of noise: the non-parametric luminescence of the medium itself (hereinafter referred to only as luminescence). The same laser beam used to pump SPDC also excites the medium and induces luminescence. This effect is intrinsic to the SPDC-based single-photon sources and constitutes a non-removable source of noise. Such noise can have severe a impact on the quality of the generated single-photon states. In fact, high-fidelity single photon sources are a crucial prerequisite for the successful implementation of a large number of quantum-optics experiments. In quantum cryptography, for instance, deteriorated single-photon states can even lead to security breach \cite{Brassard2000attack}. Similarly in quantum metrology, a low fidelity of input states of light has a negative influence on measurement precision.

In this paper we address the problem of luminescence-induced noise by analyzing time-resolved luminescence of \textbeta-BaB$_2$O$_4$ (BBO) crystals (see typical case in Fig \ref{fig:streak}). BBO is a prominent material often used in non-linear optics for its suitable properties, especially in UV region \cite{Chen1985,Bhar1989}. BBO has been used to build optical parametric amplifiers, second and third harmonics generators or tunable sources of polarization entangled photon pairs. Luminescence in BBO material has been previously discussed in some papers \cite{Sangeeta2003,Sangeeta2004,Reddy2012}. These publications however focus on thermally stimulated luminescence \cite{Sangeeta2003,Sangeeta2004} or impurities in BBO powders \cite{Reddy2012} and do not analyse luminescence in the context of noise in single photon generation. Considering the importance of single photon sources and the significant impact the presence of additional luminescence may have on their quality, we believe that our study is of sizeable importance for future development of SPDC-based single-photon sources.

The paper is organised as follows: a detailed description of our experimental setup is given in Sec. \ref{sec:setup}. Then in Sec. \ref{sec:results} we analyse spectral and temporal properties of luminescence as well as its intensity. In Section \ref{sec:results} we develop a quantitative model for the estimation of the impact of luminescence on generated single-photon states. We finally conclude in Sec. \ref{sec:concl}.

\section{Experimental setup}
\label{sec:setup}
\begin{figure}
\includegraphics[scale=1]{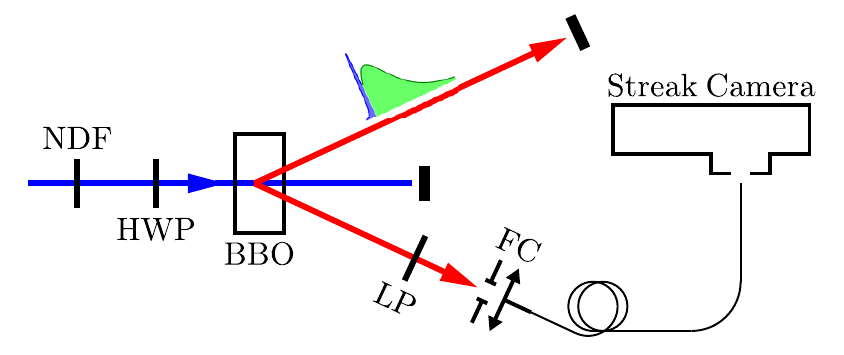}
\caption{\label{fig:setup} (color online) Scheme of the experimental setup. Optical components are denoted as follows: NDF -- neutral density filter, HWP -- half-wave plate, BBO -- BBO crystal, LP -- linear polariser, FC -- fibre coupler.}
\end{figure}
Our experimental setup (see Fig.~\ref{fig:setup}) is based on a tunable optical parametric amplifier (OPA) powered by an amplified femtosecond (fs) Ti:Sapphire laser system (Coherent Legend/Opera), which is used as the pump beam. Central wavelengths of the OPA-emitted pulses are ranging from 240 to \SI{300}{\nano\metre}. The typical OPA output power is about \SI{100}{\milli\watt} at \SI{1}{\kilo\hertz} repetition rate and the pulse duration is about \SI{200}{\femto\second}.

In the first step, we subject the OPA output beam to a tunable neutral density filter (NDF) and a half-wave plate (HWP) allowing us to set required power and polarization. After this preparation stage, the laser beam impinges on a BBO crystal, where it induces both luminescence and SPDC. Light emerging from the crystal is collected by a fiber coupler equipped with a linear polariser (LP), allowing us to perform a polarization analysis of the signal. The BBO crystal is mounted on a rotation stage positioned so that frequency-degenerate type-I SPDC process is emitted towards the fiber coupler in the different phase-matching conditions.

The signal collected by the fiber coupler is subsequently transferred by a multi-mode optical fiber to the entrance of a Czerny-Turner spectrograph, where it gets spectrally separated in the horizontal direction. Finally, in a streak camera (Hamamatsu C10910-01), the light pulses get also time-separated in the vertical direction thus creating an image with spectral resolution along the horizontal axis and time resolution along the vertical one. Typical recording of the signal is depicted in Fig \ref{fig:streak}. The crystal used throughout this experiment was manufactured by Ekspla (cut angle 48°, dimensions $8\times 8\times 5$\,mm$^3$).

\section{Results}
\label{sec:results}
Using the experimental setup described in the previous section, we acquired data for various settings of pump-pulse central wavelength, power and polarization. In this section, we provide structured presentation of the observed results.

\subsection{Spectral properties}
\begin{figure}
\includegraphics[width=8.5cm]{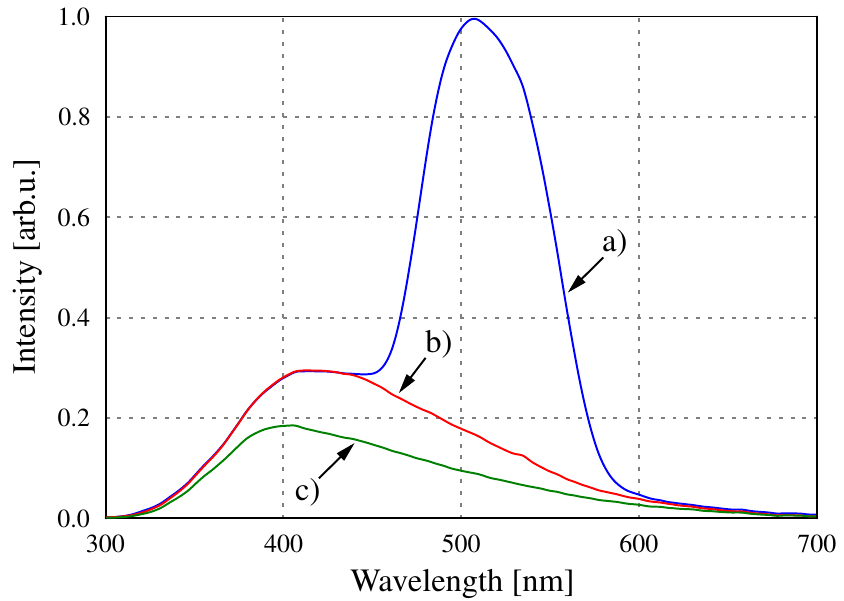}
\caption{\label{fig:spectral1} (color online) Spectral properties of both SPDC and luminescence signal: a) both SPDC and luminescence are emitted, b) SPDC is turned off by rotating the half-wave plate (HWP), c) SPDC is blocked by linear polariser (LP). See text for details.}
\end{figure}
First of all, we evaluated spectral (time-integrated) properties of both SPDC generated photons and luminescence. Typical measurement outcome is shown in Fig.~\ref{fig:spectral1}. In this figure, we show the spectra of our signal in three distinct configurations, all measured with the pump wavelength set to \SI{267}{\nano\metre}. First, we set the pump-beam polarization to satisfy phase matching conditions. In this regime, we observe a double peak blob of combined SPDC and luminescence signal. In the second configuration, we set orthogonal pump beam polarization (by rotating HWP) so that the phase-matching conditions are no longer satisfied. As expected, the SPDC is turned off and only the luminescence signal remains. Finally, we reset the pump beam polarization to satisfy the phase matching conditions and inserted a linear polarizer (LP) in front of the fiber coupler to filter out the linearly polarized SPDC. Since the luminescence photons have mixed polarization states, we still observe about one half of the luminescence passing through the linear polarizer. This set of three measurements allows us to identify spectra belonging to the SPDC and luminescence respectively. We also conclude that there is a non-negligible spectral overlap between the two.

\begin{figure}
\includegraphics[width=8.5cm]{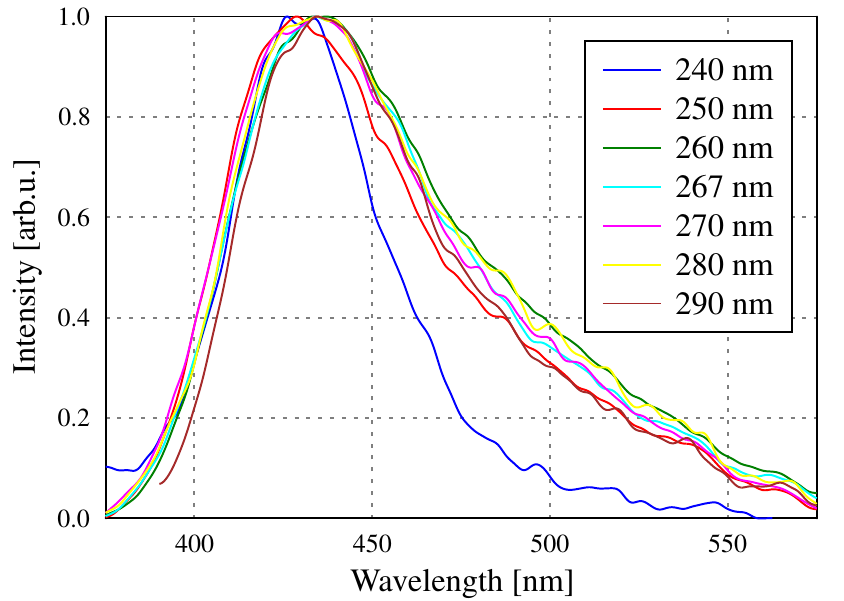}
\caption{\label{fig:spectral2} (color online) Renormalized spectra of luminescence as function of central pump wavelength.}
\end{figure}
As stated above, the BBO crystal is always rotated to satisfy degenerate phase-matching conditions. Therefore the SPDC central wavelength doubles that of the pump beam. On the other hand, luminescence spectra as a function of the central pump wavelength are to be analyzed. We have therefore acquired luminescence spectra for various pump-beam central wavelengths. The data are presented in Fig. \ref{fig:spectral2}. Note that for the sake of better readability, the spectral intensities in the plot are normalized to their peak. Fig. \ref{fig:spectral2} shows that, apart from \SI{240}{\nano\metre}, the luminescence spectra are skewed towards longer wavelengths, do not depend on the pumping wavelength and have maxima at about \SI{430}{\nano\metre}. We conclude this analysis by stating that in the range of 240 -- \SI{290}{\nano\metre} of central pump wavelength, the luminescence spectra overlap with SPDC. The overlap is more significant when using shorter pump wavelengths and is expected to reach a maximum for about \SI{220}{\nano\metre} of pump wavelength, which is unfortunately unavailable to our OPA.


To complete the spectral analysis, we have also investigated the influence of the pump power on luminescence spectra. The pump wavelength was fixed at \SI{267}{\nano\metre} while the pump power ranged from 10 -- \SI{100}{\milli\watt}. The observed results do not indicate any shift in the luminescence spectra as a function of pump power. Moreover, as expected, the luminescence intensity scales linearly with the pump power (data not shown).

All the spectral measurements indicate that there is an unavoidable spectral overlap between luminescence and SPDC when pumping with wavelengths between 240 and \SI{290}{\nano\metre}. The luminescence will thus inevitably generate noise in the SPDC signal. There are generally two possible courses of action to remove this luminescence-induced noise from the SPDC spectra. The first option is to shift the pumping wavelength above \SI{290}{\nano\metre} so that the luminescence and SPDC have only a negligible spectral overlap. This approach would however shift the SPDC central wavelength. The second option is to use edge or interference filters to cut out as much luminescence as possible. This way, one can certainly reduce the amount of noise in SPDC signal, but only to some extent given by their spectral overlap. Also note that typical interference filters have non-unit transmissivity even in their transmission maxima and therefore introduce signal losses.

We conclude this subsection stating that simple spectral filtering is not sufficient to completely remove the luminescence noise from SPDC signal and can only be used to correct this issue partially and usually also at the expenses of signal losses.

\subsection{Temporal properties}
\begin{figure}
\includegraphics[width=8.5cm]{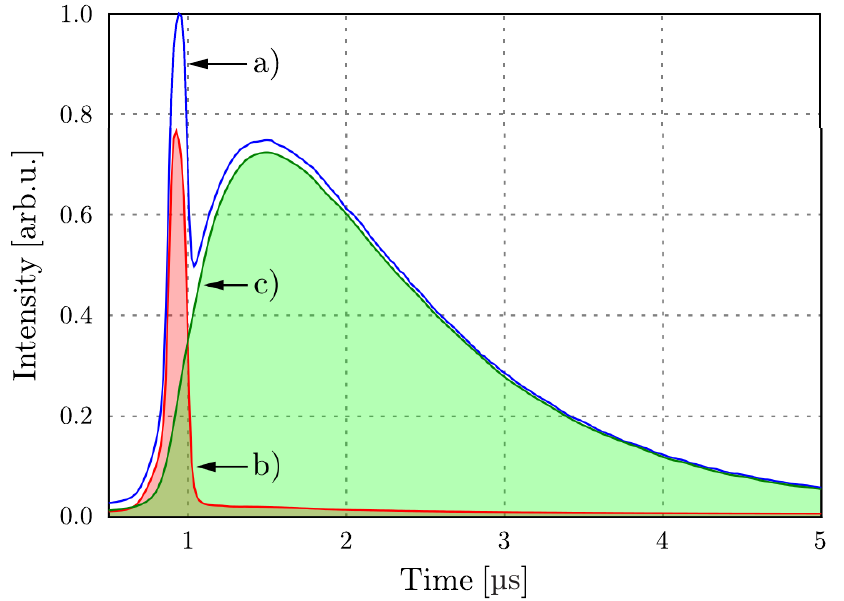}
\caption{\label{fig:time1} (color online) Temporal profile of combined SPDC and luminescence emission (a). Notice the sharp peak corresponding to the instantaneous process of SPDC (b) and the subsequent exponential decay of luminescence (c). The measured time duration of the SPDC coincides with the instrument response function (IRF) at the given streak-camera settings}
\end{figure}
The photons originating from the SPDC process are generated almost instantaneously as the pump pulse propagates through the BBO crystal, since they are described by electron transitions to virtual levels. Luminescence, on the other hand, has its typical exponential decay due to a life-time of electrons on excited energy levels. Time-resolved spectroscopy allows us to monitor this aspect of our signal as well. To separate SPDC signal from luminescence, we have selected corresponding regions of interest as schematically illustrated in Fig. \ref{fig:streak}. By using this procedure, we are able to depict typical temporal profiles of combined and separate SPDC and luminescence emissions integrated over the spectrum (see Fig. \ref{fig:time1}).

\begin{figure}
\includegraphics[width=8.5cm]{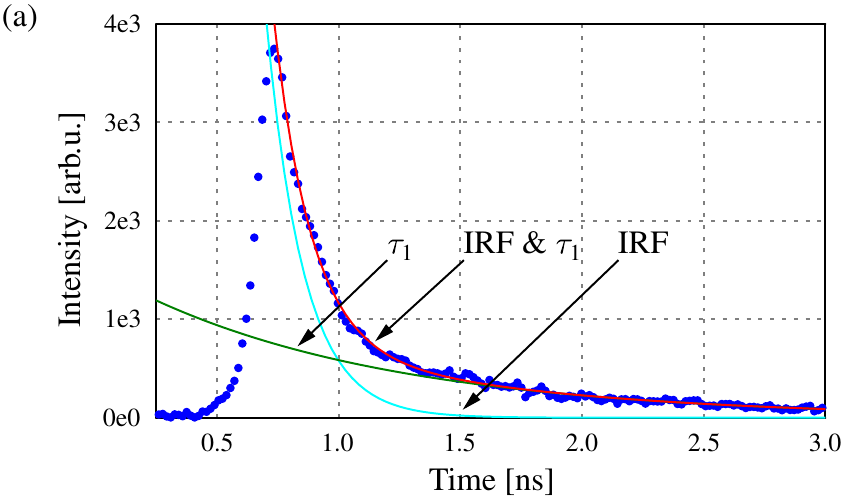}
\includegraphics[width=8.5cm]{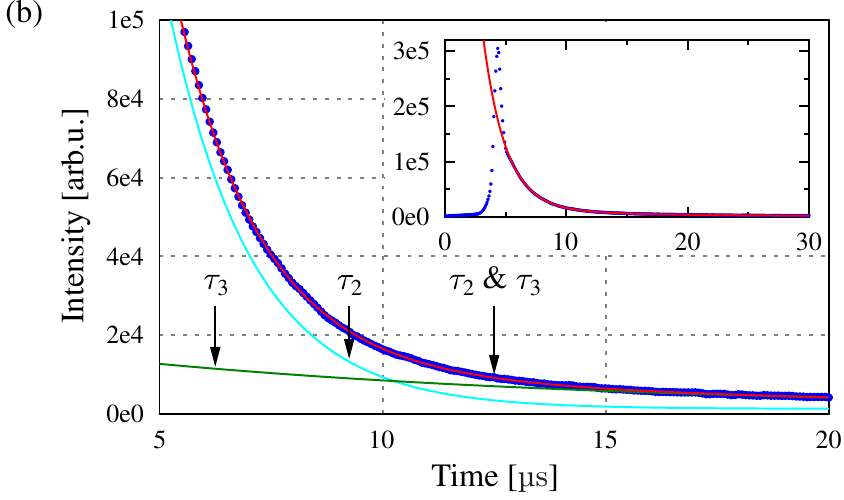}
\caption{\label{fig:time2} (color online) Exponential fit of data measured (a) in a short time-window showing a fast decay process $\tau_1=0.73$\,ns together with the instantaneous SPDC emission having the time duration of instrument response function IRF ($0.15$\,ns), (b) in longer time-window showing slower processes $\tau_2 = 1.85$\,\si{\micro\second} and $\tau_3 = 9.95$\,\si{\micro\second}. The symbol ``\&'' stands for two-exponential fit.}
\end{figure}
\begin{figure}
\includegraphics[width=8.5cm]{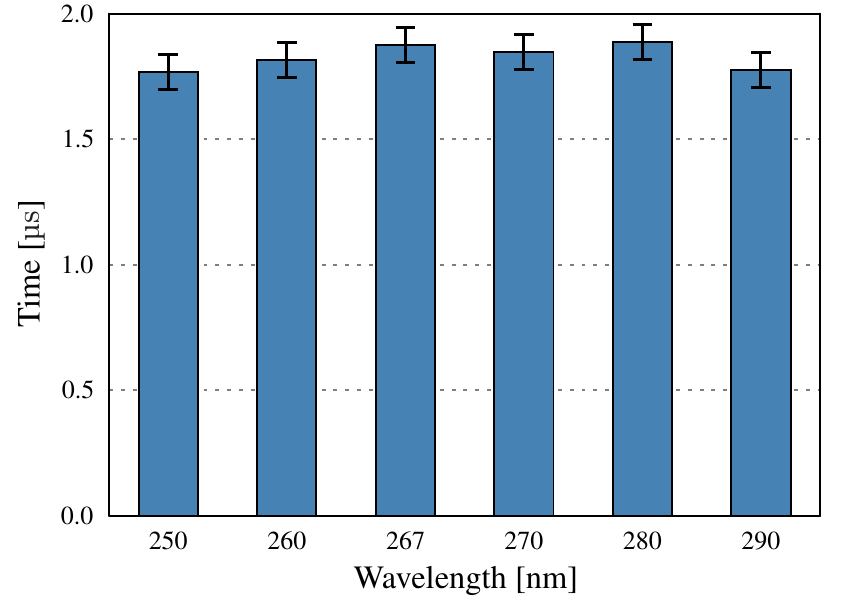}
\caption{\label{fig:time3} (color online) Decay time $\tau_2$ observed for various wavelengths of the pump beam. Note that decay times $\tau_1$ and $\tau_3$ exhibit the same trends.}
\end{figure}
The temporal properties of luminescence can be characterized by its exponential decay time $\tau$ (time period in which the intensity reaches $1/\mathrm{e}$ of its initial level. We measured time-resolved spectra at different settings of the streak-camera time window, from ns to \si{\micro\second}. By applying exponential fit to the data, we have identified several distinct decay processes with different decay times (see in Fig. \ref{fig:time2}). The first decay process is a fast one with typical decay time $\tau_1 = 0.73$\,ns. The second and third processes are considerably slower having decay times $\tau_2 = $\SI{1.85}{\micro\second} and $\tau_3 = $\SI{9.95}{\micro\metre}. The values obtained for all the decay times were established by fitting data from repeated measurements with relative uncertainty of about 7\,\%.  Moreover we have established that these decay times do not depend on the pump wavelength. As an example, see Fig. \ref{fig:time3} showing the decay time $\tau_2$ as a function of pump wavelength. Similarly, no dependence of the decay times on pumping power has been observed. 

Considering the above determined facts, we can assess the potential of temporal filtration to remove the luminescence generated noise. In quantum-optics experiments, there is usually some sort of detection window used to filter real signals from noise or detector dark counts. Typical widths of such detection window are of the order of units of nanoseconds which roughly coincides with the fastest observed decay time $\tau_1$. Therefore detection window cannot serve to filter out the fastest decaying luminescence signal. On the other hand, it can, under some circumstances, be used to filter out the slower luminescence emission. Doing so would require using pulsed pumping and reducing the overall repetition rate in the experiment (to about 100\,kHz) to allow for the luminescence to extinguish before new detection window is opened. This strategy cannot be adopted with continuous pumping since it would only limit the amount of luminescence added noise, without removing it completely.

\subsection{Intensity considerations}
So far we have determined both spectral and temporal properties of luminescence in BBO crystals. The results of the two previous subsections indicate that some reduction of luminescence induced noise can be achieved using spectral and temporal filtering. Complete removal of luminescence noise is however impossible. In this subsection we quantify the influence of luminescence on single-photon-source quality by analyzing a typical heralded single-photon source.

In order to study such model, let us consider typical parameters of heralded single photon sources: rate of single photon generation $R_S \approx 100\,$kHz (number of signal photons detected per 1\,s) and detection window $t_w \approx 10\,$ns (window in which we expect the signal photon to appear). These values are based on recent experiments in the field of linear-optical quantum information processing \cite{Lemr2011cphase,Lemr2012multi,Halenkova2012detector,Bartkiewicz2013attack}. Note that for the purposes of subsequent quantitative model, we have considered a pump wavelength of 267\,nm (fourth harmonics of Nd-YAG laser) and ideal binary detectors with unity quantum efficiency. The probability of generating a pair of SPDC photons can now be expressed as $P_S = R_S t_w \approx 1\times 10^{-3}$. Under these assumptions, we can readily neglect simultaneous generation of two SPDC photon pairs that appear with probability of $P_S^2 \approx 1\times 10^{-6}\,$. Similarly, we can define luminescence-photon rate $R_L$ and luminescence probability $P_L = R_L t_w$ to be used in forthcoming calculations. Assuming that the luminescence rate is of the same order as the SPDC rate or lower, we can neglect the probability of simultaneous generation of two luminescence photons within one detection window.

Ideally, a perfect heralded single photon source yields a signal photon in the pure Fock state $|1\rangle\langle 1|$ every time the idler photon is detected. Due to luminescence, such a source will however yield a mixed state in the form of
\begin{equation}
\label{eq:rho}
 \hat\rho = \frac{1}{N} \left( p_0 |0\rangle\langle 0| + p_1 |1\rangle\langle 1| + p_2 |2\rangle\langle 2|\right),
\end{equation}

where the normalization $N$ will be calculated later. The vacuum term $|0\rangle\langle 0|$ appears due to `false heralding' by a luminescence photon in the idler mode instead of genuine SPDC idler photon. Since luminescence photons do not originate in pairs, the signal mode is in vacuum state even if the detector in idler mode clicks. Considering that this situation happens every time a luminescence photon is detected not being accompanied by SPDC photons within one detection window, we obtain
$$
p_0 = P_L (1-P_S).
$$
The only desired term $|1\rangle\langle 1|$ appears with a success probability $p_1$ corresponding to the SPDC photon being generated in the signal mode without the luminescence photon
$$
p_1 = P_S (1-P_L).
$$

And finally, one observes the state $|2\rangle\langle 2|$ if both the SPDC and luminescence photon are emitted in the signal mode
$$
p_2 = P_S P_L.
$$
Note that $p_0 + p_1 + p_2 \neq 1$ since we have not taken into account the remaining case of neither SPDC nor luminescence photons being emitted. This case corresponds to the source not heralding signal photon presence and thus it does not contribute the effective outgoing state, but it only reduces the success probability. In order to correctly normalize the density matrix $\hat\rho$, we set
$$
N = p_0 + p_1 + p_2 = P_S(1-P_L)+P_L.
$$

The quality of generated quantum state is typically expressed by the fidelity $F$, which in our case takes the form
\begin{equation}
\label{eq:fidelity}
F = \frac{p_1}{p_0 + p_1 + p_2} = \frac{P_S (1-P_L)}{P_S (1-P_L) + P_L}.
\end{equation}

\begin{table}[!t!]
\begin{ruledtabular}
\begin{tabular}{l|cccc}
Situation & $C_S$ & $C_L$ & SNR & $F$ \\
(i) no filtering & 2.225 & 1.343 & 1.657 & 0.624\\
(ii) spectral filtering & 2.225 & 0.489 & 4.450 & 0.820 \\
(iii) spectral \& time filter. & 2.225 & 0.024 & 96.572 & 0.990
\end{tabular}
\end{ruledtabular}
\caption{\label{tab:fidelity} Measured values of SPDC and luminescence counts (in units of $1.10^{10}$) under various sets of filtering. Resulting calculation of SNR and fidelity $F$ is also presented.}
\end{table}
Time-resolved spectroscopy allows us to separate SPDC and luminescence signal by performing intensity integration over suitable regions of interest. This way, we determine the SPDC-to-luminescence count ratio, or signal-to-noise ratio
\begin{equation}
\label{eq:snr}
\mathrm{SNR} = \frac{C_S}{C_L} = \frac{R_S}{R_L},
\end{equation}
where $C_S$ and $C_L$ denote SPDC and luminescence photon counts respectively. It also follows from Eq. (\ref{eq:snr}) that realistic detector efficiency $\eta$ will not affect the resulting SNR since both $R_S$ and $R_L$ will be rescaled by the same factor $\eta$. The fidelity can be now expressed using SNR in the form of
$$
F = \frac{\mathrm{SNR}-t_wR_S}{1+\mathrm{SNR}-t_wR_S} \approx \frac{\mathrm{SNR}}{1+\mathrm{SNR}},
$$
where the last approximation holds for low generation probability.

We have calculated the expected value of fidelity for several different scenarios: (i) no spectral or time filtering of luminescence, (ii) spectral filtering only by a long-pass edge filter with cut-off at 460\,nm (e.g. continuous pump) and (iii) both spectral and temporal filtering. For all calculations, we have assumed the above mentioned typical values of $t_w$ and $R_S$ while using experimentally acquired values of SNR. Values are summarized in Table \ref{tab:fidelity}. As we can see, without simultaneous spectral and temporal-filtering, the fidelity is significantly reduced. Only when both the spectral and temporal-filtering are applied, the fidelity approaches its theoretical limit $F=1$. Spectral filtering alone can achieve about half of this improvement.

\section{Conclusions}
\label{sec:concl}
In this paper we have characterized luminescence noise in SPDC-based single-photon sources using BBO crystals. We have observed that SPDC and luminescence spectra at least partially overlap when the pump wavelength is set in the range 240 -- 290\,nm. This fact, together with mixed polarization state of luminescence photons, leads to the impossibility of a complete purification of the single photon source using spectral filtering only. In the next step, we have measured temporal characteristics of the luminescence noise. Data indicate that there are faster (ns) and slower (\si{\micro\second}) luminescence decays. At the expenses of a lower experiment repetition rate, the slower luminescence signal can be removed by a proper temporal gating. The faster process however remains an issue for typical gating windows (about 10\,ns). The last section of our study provides a quantitative model of a typical experimental configuration. We have defined the fidelity of a single-photon source and showed that only precise spectral and time-filtering allows for quasi-complete removal of the luminescence noise. In our model, spectral filtering alone allows for fidelity to exceed $0.8$, while entirely without filtering the fidelity reaches a value of about $0.6$. Note that the luminescence spectra of various BBO crystals may slightly vary depending on impurities they contain \cite{Reddy2012}.

\section*{Acknowledgements}
The authors thank Jára Cimrman for helpful suggestions. Support by projects P205/12/0382 of GA \v{C}R and projects CZ.1.05/2.1.00/03.0058 and CZ.1.07/2.3.00/20.0058 of M\v{S}MT \v{C}R are acknowledged. RM thanks also to IGA (Nos. PrF\_2013\_006 and PrF\_2014\_005). The work was also supported by MIUR under the grant agreement FIRB
``LiCHIS'' - RBFR10YQ3H. ML, AA and MB thank Luca Nardo for fruitful discussions.


\begin{thebibliography}{99}
\bibitem{Nielsen_QCQI}  M.A. Nielsen and I.L. Chuang,
{\it Quantum Computation and Quantum Information}
,   Cambridge University Press, Cambridge, 2000.

\bibitem{Zeilinger_QIP}  D.~Bouwmeester, A.~Ekert, A.~Zeilinger:  {\it The Physics of Quantum Information},   Springer, Heidelberg, 2001.

\bibitem{Halenkova2012detector} E.~Halenkov{\'a}, A.~\v{C}ernoch, K.~Lemr, J.~Soubusta, and S.~Drusov{\'a},   Appl. Opt. {\bf 51}  (4),  474--478  (2012).

\bibitem{Bartuskova2007cloning}
Bart{\accent23 u}\v{s}kov\'{a},~L.,  Du\v{s}ek,~M.,
\v{C}ernoch,~A., Soubusta,~J., and Fiur\'a\v{s}ek,~J. ,
 \textit{Phys. Rev.
Lett.}~\textbf{99}, 120505 (2007).

\bibitem{Nagali2010cloning} E.~Nagali, D.~Giovannini, L.~Marrucci, S.~Slussarenko, E.~Santamato, and F.~Sciarrino, Phys. Rev. Lett. {\bf 105}, 073602 (2010).

\bibitem{bib:fiurasek:multimode_fockstates}
J.~Fiur{\'a}\ifmmode~\check{s}\else {\v{s}}\fi{}ek, S.~Massar, N.~J. Cerf,
Phys.~Rev.~A \textbf{68}, 042325 (2003).

\bibitem{bib:zavatta:single_add}
A.~Zavatta, S.~Viciani, M.~Bellini,
Science \textbf{306}, 660 (2004).

\bibitem{bib:mitchell:super}
M.~Mitchell, J.~Lundeen, A.~Steinberg, Nature (London) \textbf{429}, 161 (2004).

\bibitem{Lemr2008engineering}
K.~Lemr and J.~Fiur{\'a}\v{s}ek,
Phys. Rev. A {\bf 77}  (2),  023802
  (2008).

\bibitem{Giovannetti2011}V. Giovannetti, S. Lloyd, L. Maccone,
 Nat. Photonics \textbf{5} 222--229 (2011).

\bibitem{Boto2000}A. N. Boto, P. Kok, D. S. Abrams, S. L. Braunstein, C. P. Williams, and J. P. Dowling,
 Phys. Rev. Lett. \textbf{85}, 2733--2736 (2000).


\bibitem{Pittman1995}T. B. Pittman, Y. H. Shih, D. V. Strekalov, and A. V.
Sergienko,
 Phys. Rev. A 52, R3429-R3432 (1995).


\bibitem{Mandel1995}L. Mandel and E. Wolf, {\it Optical Coherence and Quantum Optics}, Cambridge University Press, New York, NY.

\bibitem{Michler2000}P. Michler, A. Kiraz, C. Becher, W. V. Schoenfeld, P. M. Petroff, L. Zhang, E. Hu, A. Imamoglu,
 Science \textbf{290} 2282--2285 (2000).

\bibitem{He2013semiconductor} Y.-M.~He, Y.~He, Y.-J.~Wei, D.~Wu, M.~Atatüre, C.~Schneider, S.~Höfling, M.~Kamp, C.-Y.~Lu, and J.-W.~Pan, Nature Nanotechnology {\bf 8}, 213--217 (2013).




\bibitem{bib:bouwmeester:teleport}
D.~Bouwmeester, J.~Pan, K.~Mattle, M.~Eibl, H.~Weinfurter, A.~Zeilinger,
Nature (London) \textbf{390}, 575 (1997).

\bibitem{Kiesel2005cphase}
N.~Kiesel, C.~Schmid, U.~Weber, R.~Ursin, and H.~Weinfurter, Phys. Rev. Lett. {\bf 95}, 210505 (2005).

\bibitem{OBrien2007comp}
J.\,L.\ O'Brien, Science {\bf 318}, 1567 (2007).

\bibitem{Lemr2011cphase}
K.~Lemr, A.~\v{C}ernoch, J.~Soubusta, K.~Kieling, J.~Eisert, and M.~Du\v{s}ek,
  Phys. Rev. Lett. {\bf 106}  (1),  13602  (2011).

\bibitem{Micuda2012ampl} M.~Mičuda, I.~Straka, M.~Miková, M.~Dušek, N.~J.~Cerf, J.~Fiurášek, and M.~Ježek,  Phys. Rev. Lett. {\bf 109}, 180503 (2012).


\bibitem{Vitelli2013fusion} C.~Vitelli, N.~Spagnolo, L.~Aparo, F.~Sciarrino, E.~Santamato, L.~Marrucci, Nat. Photon. \textbf{7}, 521 (2013).

\bibitem{Hong1987}C. K. Hong, Z. Y. Ou, and L. Mandel,
Phys. Rev. Lett. \textbf{59} 2044--2046 (1987).

\bibitem{Hamar2010}M. Hamar, J. Perina Jr., O. Haderka, V. Michalek,
 Phys. Rev. A	\textbf{81}	43827 (2010).


\bibitem{BB84}
C.~Bennett and G.~Brassard, in: {\it
Proceedings of the IEEE International Conference on
Computers, Systems, and Signal Processing}  (IEEE, New York,
1984), p. 175.

\bibitem{E91}
A.~Ekert, \prl~{\bf 67}, 661 (1991).

\bibitem{gisin2002crypto}
N.~Gisin, G.~Ribordy, W.~Tittel, and H.~Zbinden, Rev. Mod. Phys. {\bf 74}, 145 (2002).

\bibitem{R04}
J.~Renes, \pra~{\bf 70}, 052314 (2004).



\bibitem{Lemr2012multi} K.~Lemr, K.~Bartkiewicz, A.~\v{C}ernoch, J.~Soubusta, and A.~Miranowicz, Phys. Rev. A {\bf 85}, 050307(R) (2012).

\bibitem{Micuda2013gate} M.~Mičuda, M.~Sedlák, I.~Straka, M.~Miková, M.~Dušek, M.~Ježek, and J.~Fiurášek, Phys. Rev. Lett. {\bf 111}, 160407 (2013).

\bibitem{Bartkiewicz2013attack} K.~Bartkiewicz, K.~Lemr, A.~\v{C}ernoch, J.~Soubusta, and A.~Miranowicz, \prl{\textbf{110}}, 173601 (2013).


\bibitem{Burnham1970source} D.~C.~Burnham and D.~L.~Weinberg, Phys. Rev. Lett. {\bf 25}, 84 (1970).

\bibitem{Chen2006source} J.~Chen, A.-J.~Pearlman, A.~Ling, J.~Fan, and A.-A.~Migdall, Opt. Express {\bf 17}, 6727--6740 (2006).

\bibitem{Walther2007source} P.~Walther, M.~Aspelmeyer, and A.~Zeilinger, Phys. Rev. A {\bf 75}, 012313 (2007).

\bibitem{Hunault2010source} M.~Hunault, H.~Takesue, O.~Tadanaga, Y.~Nishida, and M.~Asobe, Opt. Lett. {\bf 35}, 1239--1241 (2010).


\bibitem{perina2013herald} J.~Peřina, jr., O.~Haderka, V.~Michálek, Opt. Express {\bf 21}, 19389 (2013)

\bibitem{lamperti2014herald} M.~Lamperti, A.~Allevi, M.~Bondani, R.~Machulka, V.~Michálek, O.~Haderka, J.~Peřina, jr., J. Opt. Soc. Am. B {\bf 31}, 20--25 (2014).

\bibitem{Brassard2000attack} G.~Brassard, N.~Lütkenhaus, T.~Mor, and B.-C.~Sanders, Phys. Rev. Lett. {\bf 85}, 1330 (2000).

\bibitem{Chen1985}C. Chen, B. Wu, A. Jiang, and G. You,
Sci. Sin. Ser. B \textbf{28}, 235--243 (1985).

\bibitem{Bhar1989}G. C. Bhar, S. Das, and U. Chatterjee,
Appl. Opt. \textbf{28} 202--204 (1989).

\bibitem{Sangeeta2003}Sangeeta, S.C. Sabharwal,
 J. Lumin. \textbf{104} 267--271 (2003).

\bibitem{Sangeeta2004}Sangeeta, S.C. Sabharwal,
J. Lumin. \textbf{109} 69--74 (2004).

\bibitem{Reddy2012}Ch. Venkata Reddy, Ch.Rama Krishna, T. Raghavendra Rao, D.V.Sathish, P.S.Rao, R.V.S.S.N. Ravikumar,
 J. Lumin. \textbf{132} 2325--2329 (2012).
\end{thebibliography}
\end{document}